\UseRawInputEncoding
\documentclass[prd,showkeys,onecolumn,superscriptaddress]{revtex4}
\usepackage{graphicx}
\usepackage{amsmath}

\def\gap{\;\rlap{\lower 2.5pt
 \hbox{$\sim$}}\raise 1.5pt\hbox{$>$}\;}
\def\lap{\;\rlap{\lower 2.5pt
   \hbox{$\sim$}}\raise 1.5pt\hbox{$<$}\;}
\def\gsim{\;\rlap{\lower 2.5pt
 \hbox{$\sim$}}\raise 1.5pt\hbox{$>$}\;}
\def\lsim{\;\rlap{\lower 2.5pt
   \hbox{$\sim$}}\raise 1.5pt\hbox{$<$}\;}

\def\spose#1{\hbox to 0pt{#1\hss}}
\def\lta{\mathrel{\spose{\lower 3pt\hbox{$\mathchar''218$}}
     \raise 2.0pt\hbox{$\mathchar''13C$}}}
\def\gta{\mathrel{\spose{\lower 3pt\hbox{$\mathchar''218$}}
     \raise 2.0pt\hbox{$\mathchar''13E$}}}
\newcommand{\be}{\begin{equation}}
\newcommand{\ee}{\end{equation}}

\newcommand{\ls}{\mathrel{\raise1.16pt\hbox{$<$}\kern-7.0pt %  <
\lower3.06pt\hbox{{$\scriptstyle \sim$}}}}         %  ~
\newcommand{\gs}{\mathrel{\raise1.16pt\hbox{$>$}\kern-7.0pt %  >
\lower3.06pt\hbox{{$\scriptstyle \sim$}}}}         %  ~

\long\def\comment#1{}

\def\fun#1#2{\lower3.6pt\vbox{\baselineskip0pt\lineskip.9pt
  \ialign{$\mathsurround=0pt#1\hfil##\hfil$\crcr#2\crcr\sim\crcr}}}
\def\lap{\mathrel{\mathpalette\fun <}}
\def\gap{\mathrel{\mathpalette\fun >}}
\newcommand{\ba}{\begin{eqnarray}}
\newcommand{\ea}{\end{eqnarray}}

\begin{document}
\bibliographystyle{apsrev.bst}
\title{Weak cosmic censorship conjecture and black hole shadow for black hole with generalized uncertainty principle}

\author{Meirong Tang}
\email{tangmr@gzu.edu.cn(Correspondingauthor)}
\affiliation{College of Physics, Guizhou University, Guiyang 550025, China}
%\affiliation{Key Laboratory of Particle Astrophysics,
%Institute of High Energy Physics, Chinese Academy of Sciences,
%Beijing 100049, China}
%\author{Xiaobo Gong}
%\affiliation{Yunnan Observatories, Chinese Academy of Sciences, 396 Yangfangwang, Guandu District, Kunming, 650216, China}
%\affiliation{University of Chinese Academy of Sciences, Beijing 100049, China}
%\affiliation{Key Laboratory for the Structure and Evolution of Celestial Objects, Chinese Academy of Sciences, 396 Yangfangwang, Guandu District, Kunming 650216, China}
%\author{Zhaoyi Xu}
%\affiliation{College of Physics, Guizhou University, Guiyang 550025, China}
%\affiliation{Key Laboratory of Particle Astrophysics,Institute of High Energy Physics, Chinese Academy of Sciences,Beijing 100049, China}
%\affiliation{University of Chinese Academy of Sciences, Beijing 100049, China}
%\affiliation{National Astronomical
%Observatories of China, Chinese Academy of Sciences, Beijing
%100012, China}

\begin{abstract}

Based on string theory, loop quantum gravity, black hole physics, and other theories of quantum gravity, physicists have proposed generalized uncertainty principle (GUP) modifications. 
In this work, within the framework of GUP-gravity theory, we have successfully derived an exact solution to Einstein$'$s field equation, discuss the possibility of using EHT to test GUP and how GUP changes the weak cosmic censorship conjecture for black hole. We analyze two different ways of constructing GUP rotating black holes (Model I and Model II). Model I takes into account the modification of mass by GUP, i.e. the change of mass by quantization of space, and the resulting GUP-rotating black hole metric (\ref{18}) is similar in form to Kerr black hole. Model II takes into account the modification of the rotating black hole when GUP is an external field, where GUP acts like an electric charge, and the resulting GUP-rotating black hole metric (\ref{19}) is similar in form to Kerr-Newman black hole. 
By analyzing the shadow shape of the GUP rotating black hole, we have discovered intriguing characteristics regarding the impact of first-order and second-order momentum correction coefficients on the black hole's shadow shape. These findings will be instrumental in future GUP testing using EHT. Additionally, by incident test particle and scalar field with a rotating GUP-black hole, the weak cosmic censorship conjecture is not violated in either extreme black holes or near-extreme black hole.
\end{abstract}


\maketitle{}

\section{Introduction}
\label{intro}

In 1905, Einstein proposed his special relativity, which organically combined space and time. In 1915, Einstein extended his special relativity, which describes gravity phenomenon, in which he linked the effects of gravity to the curvature of space-time \cite{Charles W. Misner}. These insights revolutionized our understanding of the universe. General relativity has been highly successful where space-time is a continuous quantity. For example, the length of an object can be infinitely small and the measure of time is also a continuous quantity, but the continuity of space-time contradicts the Heisenberg uncertainty principle in quantum mechanics. In quantum mechanics, the mathematical form of Heisenberg uncertainty principle is $\Delta x \Delta p \geq \frac{\hbar}{2}$, which translates into the uncertainty of energy and time is $\Delta E \Delta t \geq \frac{\hbar}{2}$. We will see that in quantum mechanics, quantities are basically quantized, but $\Delta x$ and $\Delta t$ are continuous. In order to introduce the quantum nature of space-time into quantum mechanics, Snyder et al generalized the uncertainty principle, whose initial form is $\Delta x \Delta p \geq \frac{\hbar}{2}(1+\beta(\Delta p)^{2}+\beta\langle \hat{p}\rangle^{2})$ \cite{1989PhLB..216...41A}. This generalization introduces the notion of minimal observable length, which their work shows is not inconsistent with Lorentz invariance and is therefore a self-consistent physical theory. Since then, physicists have studied the generalized uncertainty principle (GUP) extensively and profoundly \cite{2015RPPh...78l6001N}.  

The physical discussion of GUP is important, but cannot be tested by current high energy physics experiments. Here, we only introduce the content of studying GUP by using black hole physics. The main problem is to construct a black hole metric that takes GUP into account and then to calculate the effect of GUP in black hole physics. The specific behaviour and properties of the GUP have been extensively studied when it is considered in the space-time of Schwarzschild black hole, where the radius of the event horizon of the black hole increases \cite{2018arXiv181005645F,2017PhLB..772..737F,2016EPJC...76..212F,2020EPJC...80..853B,2008PhRvL.101v1301D,2011PhRvD..84d4013A}. In loop quantum gravity theory, KK et al obtained the modified Schwarzschild black hole in the GUP case \cite{2007IJMPD..16.1397H,2004CQGra..21R..53A}. In addition, physicists also try to use GUP to solve the information problem of black holes, and find that if GUP is taken into account, the Hawking radiation of the modified Schwarzschild black hole or Kerr black hole will show a new property, that is, the black hole will leave a residue in the final stage of evaporation, which can relieve the information problem of black holes to some extent \cite{2015RPPh...78l6001N}. Currently, there are two main approaches to constructing black holes in GUP. First, the GUP does not change the equivalent metric structure of the Schwarzschild black hole, but only the total mass of the system, which is an essential modification of profound physical significance \cite{2020PhLB..81035830A}. Second, considering GUP as some outer field, similar to electric charge, so the modification in the Schwarzschild black hole metric will differ from the previous case by adding an $\propto \frac{1}{r^{2}}$ term to the space-time metric \cite{2021NuPhB.96415324O}. These two different constructions will deepen the understanding of the nature of GUP. 

In this paper, we have studied solutions of Einstein's field equations for rotating black holes under the GUP effect. In Section \ref{correction}, the GUP and Schwarzschild black hole corrections are introduced. In Section \ref{Kerr-like}, we study the GUP corrections to rotating black holes and obtain exact solutions of the Einstein field equations taking the GUP into account. In Section \ref{shadow}, we investigate the possibility of using black hole shadow to test GUP. In Section \ref{WCC-test}, we calculate the effect of GUP on the weak cosmic censorship conjecture. The last section \ref{summary} is devoted to a summary.

\section{Nonrotating Black hole and GUP}
\label{correction}

The introduction of GUP is closely related to the minimal observable length, and the introduction of GUP does not violate Lorentz invariance \cite{2020PhLB..81035830A}. Throughout this paper, we will use the following form of GUP:

\begin{equation}
\Delta x \Delta p\geq \frac{\hbar}{2}\left[ 1-\dfrac{\alpha l_{p}}{\hbar}\Delta p+\dfrac{\beta l_{p}^{2}}{\hbar^{2}}(\Delta{p})^{2}\right]
\label{1-GUP} 
\end{equation}

where $\alpha$ and $\beta$ are dimensionless model parameters. When both $\alpha$ and $\beta$ are zero, equation (\ref{1-GUP}) reduces to the Heisenberg uncertainty principle. Here $l_{p}$ is the Planck length. According to equation (\ref{1-GUP}), there exists a minimal observable length $\Delta x\geq (\Delta x)_{min}\approx \left(\sqrt{\beta}-\frac{\alpha}{2}\right)l_{p}$ and a maximal observable momentum $\Delta p\leq (\Delta p)_{max}\approx \frac{\alpha \hbar}{\beta l_{p}}$. These results show that it is impossible to measure lengths below $(\Delta x)_{min}$, setting a minimum limit to the detection capabilities of physics. In reference \cite{2020PhLB..81035830A}, they arrive at the modified black hole metric, whose space-time line elements are

\begin{equation}
ds^{2}=\left[1-\dfrac{2M_{1}}{r} \right]dt^{2}+\left[1-\dfrac{2M_{1}}{r} \right]^{-1}dr^{2}+r^{2}d\Omega^{2},  \quad \quad  (Model \; I)
\label{2-spacetime elements} 
\end{equation}

where $d\Omega^{2}=d\theta^{2}+\sin^{2}\theta d\phi^{2}$, the total mass of the modified black hole is

\begin{equation}
M_{1}=M\left[1-\dfrac{2\alpha}{M}+\dfrac{4\beta}{M^{2}} \right].
\label{3-modified BH total mass}
\end{equation}

In this approach to the construction of black hole, the GUP has a major effect on the mass and the value of $M$ deviates from the usual definition due to the presence of $\alpha$ and $\beta$. This leads us to a renewed understanding of the concept of mass. Due to the equivalence principle, gravitational mass is equivalent to inertial mass, so result (\ref{3-modified BH total mass}) will redefine our understanding of inertial mass.

For the second construction of the GUP black hole, we shall give a brief introduction. According to equation (\ref{1-GUP})

\begin{equation}
\Delta x\approx\dfrac{1}{2E}\left[1-\alpha E+\beta E^{2} \right],
\label{4}
\end{equation}

In the process of obtaining equation (\ref{4}), we use the standard dispersion relation $E=p$. On the other hand, the uncertainty in the photon wavelength depends on the Schwarzschild black hole radius

\begin{equation}
\Delta x=2\pi R_{s}=4\pi M,
\label{5}
\end{equation}

Considering $E\approx T$ and $T$ is the Bekenstein-Hawking classical temperature, the mathematical expression of $T$ can be obtained by combining (\ref{4}) and (\ref{5})

\begin{equation}
T=\dfrac{1}{2\beta}\left[\alpha+4\pi M-\sqrt{(\alpha+4\pi M)^{2}-4\beta} \right].
\label{6}
\end{equation}
 
In this construction method, the space-time line elements of the spherically symmetric GUP black hole are assumed to be \cite{2021NuPhB.96415324O}

\begin{equation}
ds^{2}=-f(r)dt^{2}+g^{-1}(r)dr^{2}+r^{2}d\Omega^{2},  \quad \quad (Model \; II)
\label{7}
\end{equation}

Its metric coefficients are set to

\begin{equation}
f(r)=g(r)=1-\dfrac{2M}{r}+\varepsilon\dfrac{M^{2}}{r^{2}},
\label{8}
\end{equation}

Here $\varepsilon$ is a dimensionless parameter. The Bekenstein-Hawking classical temperature corresponding to the metric (\ref{7}) is

\begin{equation}
T(\varepsilon)=\dfrac{1}{4\pi}\dfrac{df(r)}{dr}\vert_{r=r_{H}}=\dfrac{1}{2\pi M}\dfrac{\sqrt{1-\varepsilon}}{(1+\sqrt{1-\varepsilon})^{2}},
\label{9}
\end{equation}

Therefore, $T(\varepsilon)=T$, the relation between the GUP parameter $\beta$ and the dimensionless parameter $\varepsilon$ is

\begin{equation}
\dfrac{1}{\pi M}\dfrac{\sqrt{1-\varepsilon}}{(1+\sqrt{1-\varepsilon})^{2}}=\dfrac{1}{\beta}\left[\alpha+4\pi M-\sqrt{(\alpha+4\pi M)^{2}-4\beta} \right].$$$$
\varepsilon=1-\Big[\frac{\beta}{2\pi M\left[\alpha+4\pi M-\sqrt{(\alpha+4\pi M)^{2}-4\beta} \right]}\Big(1\pm\sqrt{1-\frac{4\pi M\left[\alpha+4\pi M-\sqrt{(\alpha+4\pi M)^{2}-4\beta} \right]}{\beta}}\Big)-1\Big]^{2}
\label{10}
\end{equation}

By the construction of spherically symmetric metric for the GUP black hole, it can be found that this construction treats the GUP as an external field rather than as a change of space-time itself. This is different from the first construction of the GUP black hole metric. Different structural pathways reveal different physical implications of the GUP.

\section{Kerr-like black hole spacetime}
\label{Kerr-like}

Here we will generalize the spherically symmetric space-time metric to the axisymmetric case using a transformation (i.e., the Newman-Janis method) \cite{1965JMP.....6..915N,2014PhLB..730...95A,2014PhRvD..90f4041A}. In this method, the black hole spin $a$ and angular coordinate $\theta$ are introduced by a complex coordinate transformation in the light cone coordinate system. By generalizing the various metric coefficients of the spherically symmetric case to the case involving $a$ and $\theta$, the general form of the space-time metric in the light cone coordinate system can be obtained. Then, using the standard coordinate transformation, we can obtain the space-time metric in the Boyer-Lindquist coordinate system, whose general expression is

\begin{equation}
ds^{2}=-\dfrac{\Psi}{\rho^{2}}\left(1-\dfrac{2\bar{f}}{\rho^{2}} \right)dt^{2}+\dfrac{\Psi}{\Delta}dr^{2}-\dfrac{4a\bar{f}\Psi\sin^{2}\theta}{\rho^{4}}dtd\phi+\Psi d\theta^{2}+\dfrac{\Psi \Sigma\sin^{2}\theta}{\rho^{4}}d\phi^{2},
\label{11-Boyer-Lindquist}
\end{equation}

The meanings of the symbols in this metric are:
\begin{equation}
\rho^{2}=k(r)+a^{2}\cos^{2}\theta, \quad  k(r)=h(r)\sqrt{\dfrac{f(r)}{g(r)}}=r^{2}, \quad  \bar{f}=\dfrac{1}{2}k(r)-\dfrac{1}{2}h(r)f(r)=\dfrac{1}{2}r^{2}-\dfrac{1}{2}r^{2}f(r), $$$$
\Sigma=\left(k(r)+a^{2}\right)^{2}-a^{2}\Delta(r)\sin^{2}\theta=\left(r^{2}+a^{2}\right)^{2}-a^{2}\Delta(r)\sin^{2}\theta, \quad  \Delta(r)=r^{2}f(r)+a^{2}.
\label{12}
\end{equation}

In these expressions, once the metric of spherically symmetric space-time is determined, all functions except the unknown function $\Psi$ in equation (\ref{11-Boyer-Lindquist}) can be determined. So far, two physical conditions have not been treated, namely the axisymmetric condition $G_{r\theta}=0$ and Einstein's gravitational field equations $G_{\mu\nu}=R_{\mu\nu}-\frac{1}{2}g_{\mu\nu}R=8\pi T_{\mu\nu}$. The space-time metric (\ref{11-Boyer-Lindquist}) must satisfy both conditions. In this case, the unknown function in equation (\ref{11-Boyer-Lindquist}) will satisfy the following equations:

\begin{equation}
\left(k(r)+a^{2}y^{2} \right)^{2} \left(3\dfrac{\partial\Psi}{\partial r}\dfrac{\partial\Psi}{\partial y^{2}}-2\Psi\dfrac{\partial^{2}\Psi}{\partial r \partial y^{2}} \right)=3a^{2}\dfrac{\partial k}{\partial r}\Psi^{2},
\label{13}        
\end{equation}

\begin{equation}
\Psi \left( \left(\dfrac{\partial k}{\partial r} \right)^{2}+k\left(2-\dfrac{\partial^{2}k}{\partial r^{2}} \right)-a^{2}y^{2}\left(2+\dfrac{\partial^{2}k}{\partial r^{2}} \right)\right)+ \left(k+a^{2}y^{2} \right) \left(4y^{2}\dfrac{\partial\Psi}{\partial y^{2}}-\dfrac{\partial k}{\partial r}\dfrac{\partial\Psi}{\partial r} \right)=0,
\label{14}        
\end{equation}

For the space-time metrics (\ref{2-spacetime elements}) and (\ref{7}), $f(r)=g(r)$ and $k(r)=r^{2}$, then equations (\ref{13}) and (\ref{14}) are simplified as

\begin{equation}
\left(r^{2}+a^{2}y^{2} \right)^{2} \left[3\dfrac{\partial\Psi}{\partial r}\dfrac{\partial\Psi}{\partial y^{2}}-2\Psi\dfrac{\partial^{2}\Psi}{\partial r \partial y^{2}} \right]=6a^{2}r\Psi^{2}.
\label{15}
\end{equation}

\begin{equation}
\Psi\left[4r^{2}-4a^{2}y \right]+\left(r^{2}+a^{2}y^{2} \right)\left(4y^{2}\dfrac{\partial\Psi}{\partial y^{2}}-2r\dfrac{\partial\Psi}{\partial r} \right)=0.
\label{16}
\end{equation}

By solving equations (\ref{15}) and (\ref{16}), it is found that $\Psi(r,\theta,a)$ can be written in the following form:

\begin{equation}
\Psi(r,\theta,a)=r^{2}+a^{2}\cos^{2}\theta.
\label{17}
\end{equation}

Now, we have obtained the axisymmetric form corresponding to a given spherically symmetric metric, so that once a spherically symmetric black hole metric is given, a rotating black hole solution is obtained.

(a) Model $I$

\begin{equation}
ds^{2}=-\left[1-\dfrac{2Mr-4\alpha r+\frac{8\beta}{M}r}{\rho^{2}} \right]dt^{2}+\dfrac{\rho^{2}}{\Delta}dr^{2}-\dfrac{2a\sin^{2}\theta \left[2Mr-4\alpha r+\frac{8\beta}{M}r \right]}{\rho^{2}}dtd\phi+\rho^{2}d\theta^{2}+\dfrac{\Sigma\sin^{2}\theta}{\rho^{2}}d\phi^{2},
\label{18}
\end{equation}

$ \rho^{2}=r^{2}+a^{2}\cos^{2}\theta $, $ \Sigma=\left(r^{2}+a^{2} \right)^{2}-a^{2}\Delta\sin^{2}\theta $, $ \Delta=r^{2}-2M \left(1-\dfrac{2\alpha}{M}+\dfrac{4\beta}{M^{2}} \right)r+a^{2} $.
 
This space-time metric describes the modification of the Kerr black hole when the GUP is considered as an internal property of space-time. When GUP is not taken into account, that is, $\alpha=\beta=0$, the space-time metric degenerates to a Kerr black hole.

Next, we analyze the fundamental properties of the black hole solution. When we generalize spherically symmetric GUP black holes to axisymmetric, the metric (\ref{18}) exists at least one event horizon. According to the algebraic equation $\Delta=r^{2}-2M \left(1-\frac{2\alpha}{M}+\frac{4\beta}{M^{2}} \right)r+a^{2}=0$ satisfied by the event horizon of the black hole, the solution of this equation can be obtained as $r_{\pm}=M\left[1-\frac{2\alpha}{M}+\frac{4\beta}{M^{2}} \right]\pm \sqrt{M^{2}\left(1-\frac{2\alpha}{M}+\frac{4\beta}{M^{2}} \right)^{2}-a^{2}}$, where $r_{+}$ is the event horizon and $r_{-}$ is the causal horizon. If both $r_{+}$ and $r_{-}$ have physical meaning, then $M^{2}\left(1-\frac{2\alpha}{M}+\frac{4\beta}{M^{2}} \right)^{2}-a^{2}\geq 0$ is required, which is a strong constraint on the GUP parameters $\alpha$ and $\beta$.

According to $r_{\pm}=M\left[1-\frac{2\alpha}{M}+\frac{4\beta}{M^{2}} \right]\pm \sqrt{M^{2}\left(1-\frac{2\alpha}{M}+\frac{4\beta}{M^{2}} \right)^{2}-a^{2}}$, when $\alpha=\beta=0$, $r_{\pm}=M \pm \sqrt{M^{2}-a^{2}}$, then $r_{\pm}$ degenerates to the Kerr black hole case; when $\frac{4\beta}{M^{2}}-\frac{2\alpha}{M}=0$, that is $2\beta=\alpha M$, $r_{\pm}=M \pm \sqrt{M^{2}-a^{2}}$, this is also the case of Kerr black holes, so this is a very interesting phenomenon; when the first and second terms of GUP satisfy certain conditions (i.e., $2\beta=\alpha M$), Model $I$ will be reduced to the vacuum solution case. It is easy to see from the expression of $\Delta$ that the GUP parameters $\alpha$ and $\beta$ have opposite changes on the properties of Kerr black hole.

(b) Model $II$

\begin{equation}
ds^{2}=-\left[1-\dfrac{2Mr-\varepsilon M^{2}}{\rho^{2}} \right]dt^{2}+\dfrac{\rho^{2}}{\Delta}dr^{2}-\dfrac{2a\sin^{2}\theta \left[2Mr-\varepsilon M^{2} \right]}{\rho^{2}}dtd\phi+\rho^{2}d\theta^{2}+\dfrac{\Sigma\sin^{2}\theta}{\rho^{2}}d\phi^{2},
\label{19}
\end{equation}

$ \rho^{2}=r^{2}+a^{2}\cos^{2}\theta $, $ \Sigma=\left(r^{2}+a^{2} \right)^{2}-a^{2}\Delta\sin^{2}\theta $, $ \Delta=r^{2}-2Mr+\varepsilon M^{2}+a^{2} $.

This black hole metric describes the modification of the Kerr black hole when the GUP is taken to be an external matter field. When GUP is not considered (i.e. there is no matter field), $\varepsilon=0$, then the black hole metric degenerates to the Kerr black hole.

Next, we analyze the constraints on the model parameters in the black hole solution. In order for the space-time metric (\ref{19}) to really describe a black hole, the metric must have an event horizon; mathematically, this condition can be expressed as if $\Delta=r^{2}-2Mr+\varepsilon M^{2}+a^{2}=0$ has at least one real root, that is, $r_{\pm}=M \pm \sqrt{(1-\varepsilon)M^{2}-a^{2}}$, then $\varepsilon\leq 1-\frac{a^{2}}{M^{2}}$. This is a strong constraint on the value of the GUP parameter $\varepsilon$.

From the spherically symmetric GUP black hole metrics (\ref{2-spacetime elements}) and (\ref{7}) and the axisymmetric GUP black hole metrics (\ref{18}) and (\ref{19}), it can be seen that there are obvious differences in the black hole metric constructed in different ways. In order to make (\ref{18}) have some common properties with the black hole described by (\ref{19}), we need to make them equivalent in some properties, and then find the relation between the two. 
The relationship between the parameters of model I and model II can be derived through the equivalence of Hawking temperatures in the case of spherical symmetry. However, when it comes to rotating black holes, such thermodynamic equivalence does not exist, which is a noteworthy aspect for discussion.

By comparing the black hole metrics (\ref{18}) and (\ref{19}), we find that: for the black hole metric (\ref{18}), the space-time linear element $ds^{2}$ is consistent with the Kerr black hole, except that $M$ is multiplied by a modification term $(1-\frac{2\alpha}{M}+\frac{4\beta}{M^{2}})$ determined by the GUP parameters. For the black hole metric (\ref{19}), the space-time linear element $ds^{2}$ is consistent with the Kerr-Newman black hole, and the metric (\ref{19}) can be obtained by replacing $Q$ with $\sqrt{\varepsilon}M$ in the Kerr-Newman black hole metric. Thus, we can obtain the following results: if GUP is considered as a modification to the Kerr black hole when it is an internal property of space-time, then the properties of the black hole (described by the metric (\ref{18})) are consistent with those of the Kerr black hole; if GUP is considered as a modification to the Kerr black hole when it is an external field, then the properties of the black hole (described by the metric (\ref{19})) are consistent with those of the Kerr-Newman black hole.

\section{Black hole shadow}
\label{shadow}

\subsection{Geodesics and photon spheres}
\label{geodesic}
Photons move along geodesics in the space-time background of a rotating black hole. In order to study the orbital motion of a photon, it is necessary to study the geodesic equation of the photon in the space-time of a rotating black hole. Here we will write down the main calculation process and the analysis ideas \cite{1968CMaPh..10..280C,Chandrasekhar}. In general, the Laplacian of the space-time of a rotating black hole can be expressed as

\begin{equation}
\mathcal{L}=\dfrac{1}{2}g_{\mu\nu}\dot{x}^{\mu}\dot{x}^{\nu},
\label{24}
\end{equation}

Where $g_{\mu\nu}$ is the metric coefficient matrix determined by the black hole metrics (\ref{18}) and (\ref{19}), $x^{\mu}$ is the four-dimensional space-time coordinates, and the symbol '$\cdot$' represents the derivative of affine parameter $\tau$ on the geodesic. Through the Laplacian (\ref{24}), the four-dimensional momentum of the photon can be calculated as follows:

\begin{equation}
p_{t}=\dfrac{\partial\mathcal{L}}{\partial\dot{t}}=\left[\dfrac{a^{2}\sin^{2}\theta}{\rho^{2}}-\dfrac{\Delta}{\rho^{2}} \right]\dot{t}+\left[\dfrac{a\Delta\sin^{2}\theta}{\rho^{2}}-\dfrac{a\left(a^{2}+r^{2} \right)\sin^{2}\theta}{\rho^{2}} \right]\dot{\phi},
\label{25}
\end{equation}

\begin{equation}
p_{r}=\dfrac{\partial\mathcal{L}}{\partial\dot{r}}=\dfrac{\rho^{2}}{\Delta}\dot{r}
\label{26}
\end{equation}

\begin{equation}
p_{\theta}=\dfrac{\partial\mathcal{L}}{\partial\dot{\theta}}=\rho^{2}\dot{\theta},
\label{27}
\end{equation}

\begin{equation}
p_{\phi}=\dfrac{\partial\mathcal{L}}{\partial\dot{\phi}}=\left[\dfrac{a\Delta\sin^{2}\theta}{\rho^{2}}-\dfrac{a\left(a^{2}+r^{2} \right)\sin^{2}\theta}{\rho^{2}} \right]\dot{t}+\left[\dfrac{\left(a^{2}+r^{2} \right)^{2}\sin^{2}\theta}{\rho^{2}}-\dfrac{a^{2}\Delta\sin^{4}\theta}{\rho^{2}} \right]\dot{\phi}.
\label{28}
\end{equation}

Since the black hole in question is steady-state and axisymmetric, $\mathcal{L}$ is not an explicit function of time $t$ and angular coordinate $\phi$, which makes $p_{t}$ and $p_{\phi}$ constant functions, denoted $-E$ and $L_{\phi}$, respectively. To obtain the geodesic equation for the photon, we start with the Hamilton-Jacobi equation satisfied by the photon, which has the form

\begin{equation}
-\dfrac{\partial S}{\partial \tau}=\dfrac{1}{2}g^{\mu\nu}\dfrac{\partial S}{\partial x^{\mu}}\dfrac{\partial S}{\partial x^{\nu}}.
\label{29}
\end{equation}

Here $S$ is the principal function of the Hamiltonian. If $S$ can be solved by separating variables, then the basic form of $S$ reduces to $ S=\frac{1}{2}m^{2}\sigma-Et+L_{\phi}\phi+S_{\theta}(\theta)+S_{r}(r) $. For photon geodesics, $S$ can be reduced to $ S=-Et+L_{\phi}\phi+S_{\theta}(\theta)+S_{r}(r) $. By introducing a Cartesian integration constant $\mathcal{K}$ and the functions $R$ and $H$ corresponding to $S_{r}$ and $S_{\theta}$, we can obtain the separated variable solutions of the photon geodesics, which are   

\begin{equation}
\rho^{2}\dfrac{dt}{d\tau}=E\left[ \dfrac{\left(r^{2}+a^{2}\right)\left(r^{2}+a^{2}-a\lambda \right)}{\Delta}+a\left(\lambda-a\sin^{2}\theta \right) \right],
\label{30}
\end{equation}

\begin{equation}
\left(\rho^{2}\dfrac{dr}{d\tau}\right)^{2}=R,
\label{31}
\end{equation}

\begin{equation}
\left(\rho^{2}\dfrac{d\theta}{d\tau}\right)^{2}=H,
\label{32}
\end{equation}

\begin{equation}
\rho^{2}\dfrac{d\phi}{d\tau}=E\left[ \dfrac{a\left(r^{2}+a^{2}\right)-a^{2}\lambda}{\Delta}+\dfrac{\lambda-a\sin^{2}\theta}{\sin^{2}\theta} \right].
\label{33}
\end{equation}

The functions $R$, $H$, $\lambda$ and $\eta$ are as follows:

\begin{equation}
R=E^{2}\left[\left(r^{2}+a^{2}-a\lambda \right)^{2}-\eta\Delta \right],
\label{34}
\end{equation}

\begin{equation}
H=E^{2}\left[\eta-\left(\dfrac{\lambda}{\sin\theta}-a\sin\theta \right)^{2} \right],
\label{35}
\end{equation}

\begin{equation}
\lambda=\dfrac{L_{\phi}}{E},
\label{36}
\end{equation}

\begin{equation}
\eta=\dfrac{\mathcal{K}}{E^{2}}.
\label{37}
\end{equation}

Due to the strong gravitational field and extreme properties of the black hole, there is a minimal stable orbit $r_{c}$ in its vicinity. By examining the basic properties of the function $R$, the mathematical conditions satisfied by the minimal stable orbit are

\begin{equation}
 \begin{aligned}
 &R(r_{c})=0\\
 &\dfrac{dR(r)}{dr}\vert_{r=r_{c}}=0.\\
 \end{aligned}
\label{38}
\end{equation}

By substituting equation $(\ref{34})$ into the above equations, the expression of functions $\lambda$ and $\eta$ can be obtained.

For Model $I$, the expression is

\begin{equation}
\lambda=\dfrac{1}{a}\left[2+r-\dfrac{2M}{r}\left(1-\dfrac{2\alpha}{M}+\dfrac{4\beta}{M^{2}} \right) \right]^{-1}\left(\left(r^{2}+a^{2} \right)\left(2+r-\dfrac{2M}{r}\left(1-\dfrac{2\alpha}{M}+\dfrac{4\beta}{M^{2}} \right) \right) \right. $$$$ \left.
-4\left(r^{2}-2Mr\left(1-\dfrac{2\alpha}{M}+\dfrac{4\beta}{M^{2}} \right)+a^{2} \right) \right),
\label{39}
\end{equation}

\begin{equation}
\eta=\dfrac{r^{3}}{a^{2}}\left[2+r-\dfrac{2M}{r}\left(1-\dfrac{2\alpha}{M}+\dfrac{4\beta}{M^{2}} \right) \right]^{-2}\left(8a^{2}+\dfrac{16Ma^{2}}{r^{2}}\left(1-\dfrac{2\alpha}{M}+\dfrac{4\beta}{M^{2}} \right)-r\left(r-2+\dfrac{2M}{r}\left(1-\dfrac{2\alpha}{M}+\dfrac{4\beta}{M^{2}} \right) \right)^{2} \right).
\label{40}
\end{equation}

For Model $II$, the expression is

\begin{equation}
\lambda=\dfrac{1}{a}\left[2+r-\dfrac{2M}{r} \right]^{-1}\left(\left(r^{2}+a^{2} \right)\left(2+r-\dfrac{2M}{r} \right) 
-4\left(r^{2}-2Mr+\varepsilon M^{2}+a^{2} \right) \right),
\label{41}
\end{equation}

\begin{equation}
\eta=\dfrac{r^{3}}{a^{2}}\left[2+r-\dfrac{2M}{r} \right]^{-2}\left(8a^{2}+\dfrac{16Ma^{2}}{r^{2}}-\dfrac{16\varepsilon M^{2}a^{2}}{r^{3}}-r\left(r-2+\dfrac{6M}{r}-\dfrac{4\varepsilon M^{2}}{r^{2}} \right)^{2} \right).
\label{42}
\end{equation}

By substituting equation $(\ref{38})$ into equation $(\ref{34})$, we can get the second derivative of the function $R$

\begin{equation}
\dfrac{d^{2}R}{dr^{2}}\vert_{r_{c}}=8E^{2}\left[r^{2}+\dfrac{2r\Delta \left(\Delta^{'}-r\Delta^{''} \right)}{\Delta^{'2}} \right]\vert_{r=r_{c}}.
\label{43}
\end{equation}

The stability of photon orbit can be judged by the positive or negative of $ \frac{d^{2}R}{dr^{2}}\vert_{r_{c}} $. When $ \frac{d^{2}R}{dr^{2}}\vert_{r_{c}}>0 $, the geodesic motion of the photon is unstable, and vice versa. For $ \frac{d^{2}R}{dr^{2}}\vert_{r_{c}}>0 $, the set of photon orbits is the shape of the black hole shadow. Only when function $ H\geq 0 $, photon orbit can form a photon sphere. This condition can be simplified as follows:

For Model $I$ 

\begin{equation}
\left(4r^{3}-8Mr^{2}\left(1-\dfrac{2\alpha}{M}+\dfrac{4\beta}{M^{2}} \right)+4ra^{2}-\left(r^{2}+a^{2}\cos^{2}\theta \right)\left(2r-2M\left(1-\dfrac{2\alpha}{M}+\dfrac{4\beta}{M^{2}} \right) \right) \right)^{2}\vert_{r=r_{c}}   $$$$
\leq 16a^{2}r^{2} \left(r^{2}-2Mr\left(1-\dfrac{2\alpha}{M}+\dfrac{4\beta}{M^{2}} \right)+a^{2} \right)\sin^{2}\theta\vert_{r=r_{c}}.
\label{44}
\end{equation}

For Model $II$

\begin{equation}
\left(4r^{3}-8Mr^{2}+4\varepsilon M^{2}r+4ra^{2}-\left(r^{2}+a^{2}\cos^{2}\theta \right)\left(2r-2M \right) \right)^{2}\vert_{r=r_{c}} \leq 16a^{2}r^{2} \left(r^{2}-2Mr+\varepsilon M^{2}+a^{2} \right)\sin^{2}\theta\vert_{r=r_{c}}.
\label{45}
\end{equation}

Equations $(\ref{44})$ and $(\ref{45})$ both have two positive real roots (denoted as $ r_{c_{+}} $ and $ r_{c_{-}} $, respectively), and the value range of the photon sphere is $ r_{c_{-}}\leq r_{c}\leq r_{c_{+}} $.

\subsection{Black hole shadow shape}
\label{shadow-shape}
The calculation and discussion in section \ref{geodesic} reveal that the motion of photons near the GUP black hole gives rise to the formation of a photon shape, which can be attributed to the presence of a minimum stable orbit within the spacetime of the GUP- black hole. Because the generalized uncertainty principle is taken into account, the calculation results of minimum stable orbital radius and photon sphere are different from those of Kerr black holes. 
Therefore, it is possible to test the correctness of the GUP by using black hole shadows. The research progress of black hole shadow and related calculation methods can be referred to the series of literatures e.g.\cite{2022PhR...947....1P, 2020EPJC...80..343N, 2021AnPhy.43468662A, 2023EPJC...83..679L, 2022PhLB..82636894T, 2022IJGMM..1950068J, 2023PDU....4101249K, 2023arXiv230907442O}.  Next, we will calculate the image formed by photons emitted from the photon sphere measured by observers on Earth, and examine the influence of different GUP model parameters on the shape of the black hole shadow, and then propose the possibility of using the observation of the black hole shadow and its photon ring to test it. 
%Fig 1
\begin{figure}[htbp]
  \centering
   \includegraphics[scale=0.3]{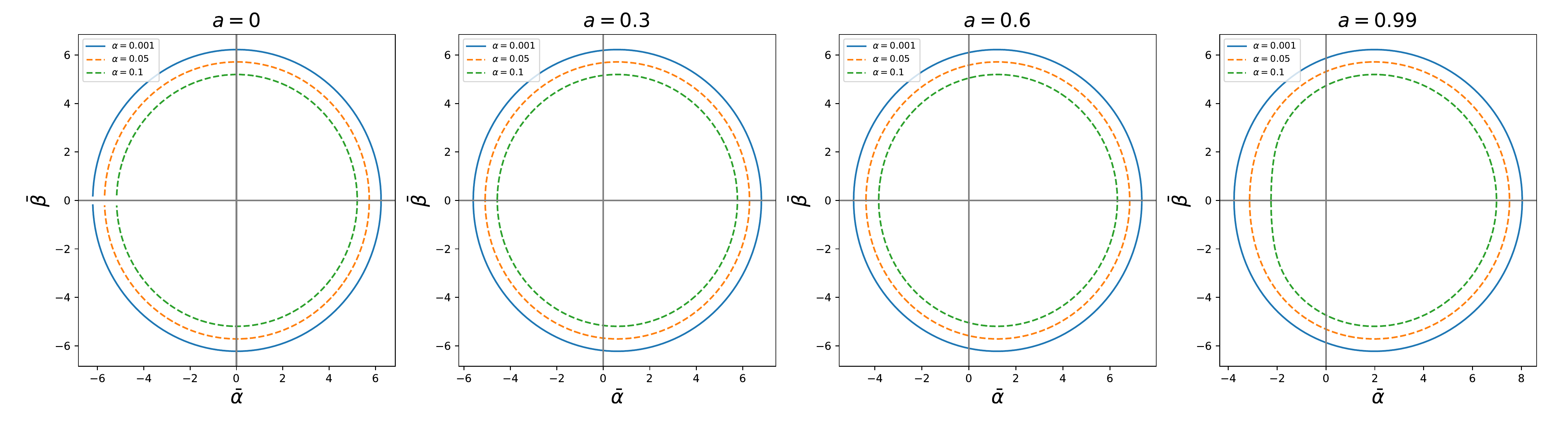}
   \includegraphics[scale=0.3]{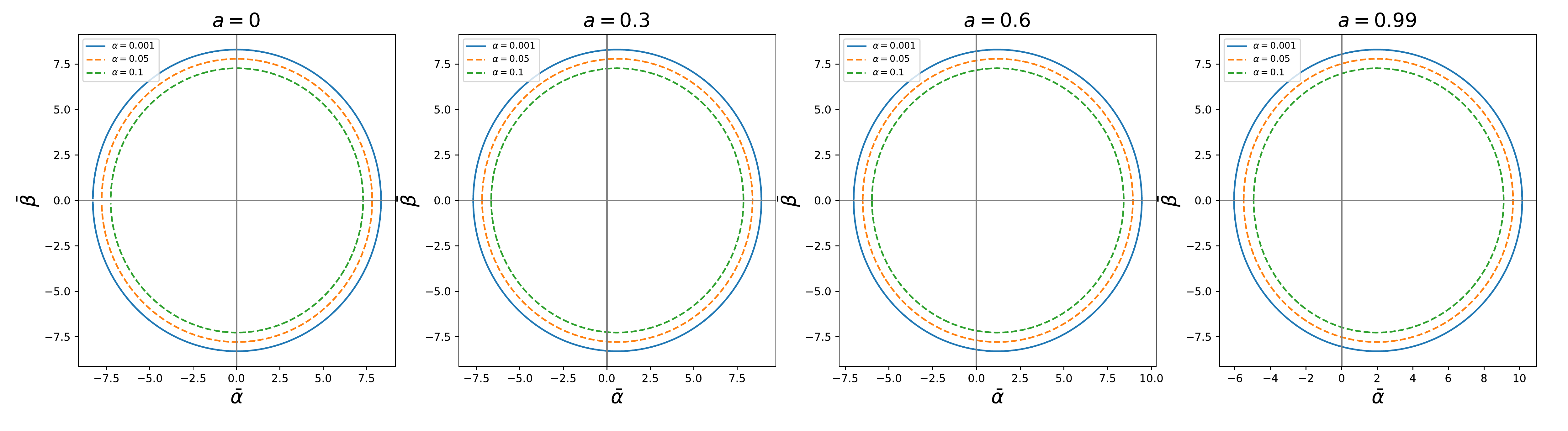}
   \includegraphics[scale=0.3]{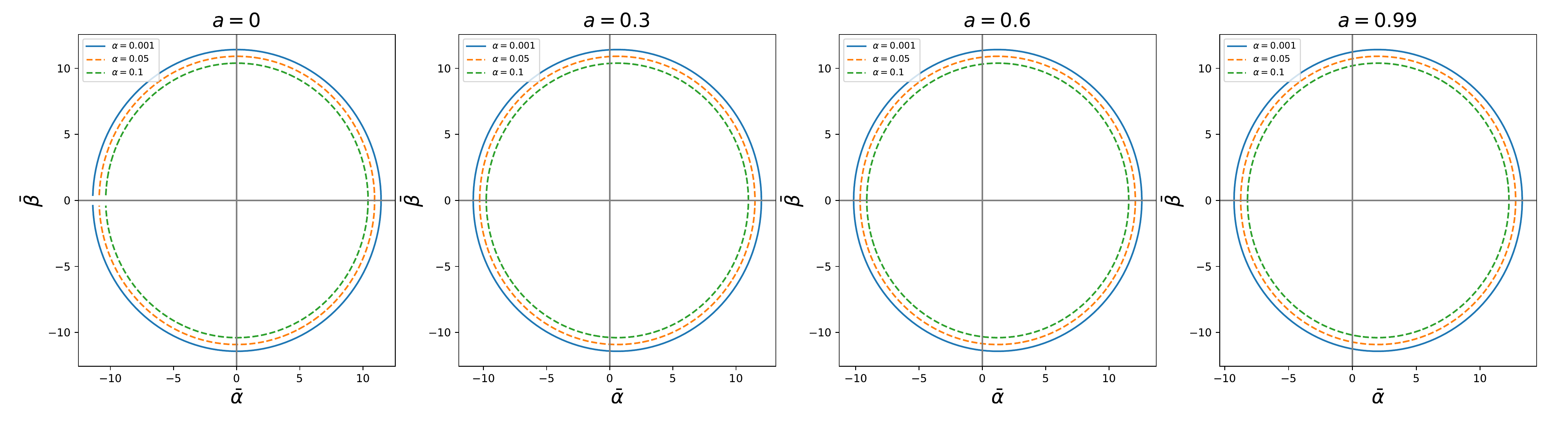}
   \caption{The shadow shape of GUP rotating black hole under different model parameters (Model I). From left to right represents the process of increasing the spin of the black hole, which is $a=0, 0.3, 0.6, 0.99$ respectively. From top to bottom represents the process of increasing first-order momentum correction coefficient, which is $\beta=0.05, 0.15, 0.3$ respectively. Curves of various colors correspond to different $\alpha$ values.}
  \label{shadow_model1_omega}
\end{figure}

%Fig 2
\begin{figure}[htbp]
  \centering
   \includegraphics[scale=0.3]{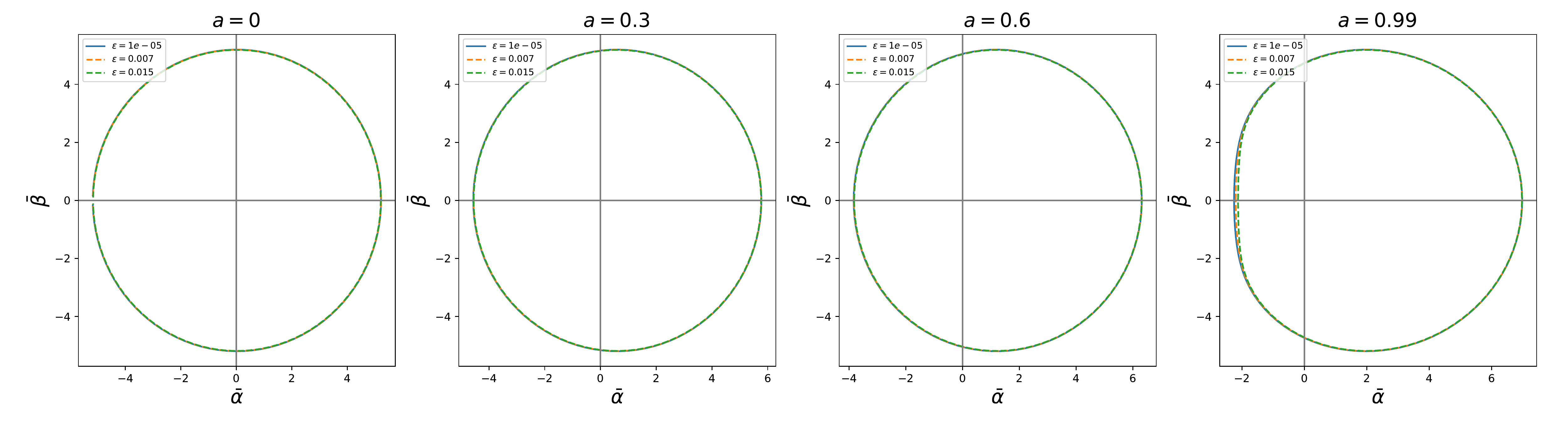}
   \caption{The shadow shape of GUP rotating black hole under different model parameters (Model II). From left to right represents the process of increasing the spin of the black hole, which is $a=0, 0.3, 0.6, 0.99$ respectively. From top to bottom represents the process of increasing mixing correction coefficient, which is $\varepsilon=1e-05, 0.007, 0.015$ respectively. Curves of various colors correspond to different $\varepsilon$ values.  }
  \label{shadow_model2}
\end{figure}

When we measure the black hole photon sphere on Earth, we generally need to approximate the model by assuming that the observer is a stationary observer at infinity, and that the observer can be approximated to a point. In the calculation of the black hole shadow, the celestial coordinate system is generally chosen, which uses two coordinates ($\bar{\alpha}, \bar{\beta}$) to describe the position of the black hole shadow in the celestial coordinate system, similar to a projection. 
According to physical understanding, the coordinates $\bar{\alpha}, \bar{\beta}$ in the celestial coordinate system should be related to the angular coordinates in the BL coordinate system, and the specific relationship is as follows
\begin{equation}
\bar{\alpha}=\mathop{\lim}_{r_{0}\rightarrow\infty}\Big[-r^{2}_{0}\sin\theta_{0}\dfrac{d\phi}{dr}\Big],$$$$
\bar{\beta}=\mathop{\lim}_{r_{0}\rightarrow\infty}\Big[r^{2}_{0}\dfrac{d\theta}{dr}\Big].
\label{shape1}
\end{equation}
If the observer is on the equatorial plane, the condition $\theta=\pi/2$is satisfied. The coordinates in the celestial coordinate system will then take the following form
\begin{equation}
\bar{\alpha}=-\lambda,$$$$
\bar{\beta}=\pm\sqrt{\eta}.
\label{shape2}
\end{equation}
For the GUP black holes discussed here, specific expressions can be obtained. For model I, its celestial coordinate expression is
\begin{equation}
\bar{\alpha}=\dfrac{-1}{a}\left[2+r-\dfrac{2M}{r}\left(1-\dfrac{2\alpha}{M}+\dfrac{4\beta}{M^{2}} \right) \right]^{-1}\left(\left(r^{2}+a^{2} \right)\left(2+r-\dfrac{2M}{r}\left(1-\dfrac{2\alpha}{M}+\dfrac{4\beta}{M^{2}} \right) \right) \right. $$$$ \left.
-4\left(r^{2}-2Mr\left(1-\dfrac{2\alpha}{M}+\dfrac{4\beta}{M^{2}} \right)+a^{2} \right) \right),$$$$
\bar{\beta}=\pm\sqrt{\dfrac{r^{3}}{a^{2}}\left[2+r-\dfrac{2M}{r}\left(1-\dfrac{2\alpha}{M}+\dfrac{4\beta}{M^{2}} \right) \right]^{-2}\left(8a^{2}+\dfrac{16Ma^{2}}{r^{2}}\left(1-\dfrac{2\alpha}{M}+\dfrac{4\beta}{M^{2}} \right)-r\left(r-2+\dfrac{2M}{r}\left(1-\dfrac{2\alpha}{M}+\dfrac{4\beta}{M^{2}} \right) \right)^{2} \right)}.
\label{shape3}
\end{equation}
For model II, its celestial coordinate expression is
\begin{equation}
\bar{\alpha}=\dfrac{-1}{a}\left[2+r-\dfrac{2M}{r} \right]^{-1}\left(\left(r^{2}+a^{2} \right)\left(2+r-\dfrac{2M}{r} \right)-4\left(r^{2}-2Mr+\varepsilon M^{2}+a^{2} \right) \right),$$$$
\bar{\beta}=\pm\sqrt{\dfrac{r^{3}}{a^{2}}\left[2+r-\dfrac{2M}{r} \right]^{-2}\left(8a^{2}+\dfrac{16Ma^{2}}{r^{2}}-\dfrac{16\varepsilon M^{2}a^{2}}{r^{3}}-r\left(r-2+\dfrac{6M}{r}-\dfrac{4\varepsilon M^{2}}{r^{2}} \right)^{2} \right)}.
\label{shape4}
\end{equation}
According to equation (\ref{shape4}), it can be found that the celestial coordinates ($\bar{\alpha}, \bar{\beta}$) are functions of the GUP parameter and the spin of the black hole. 
By using $\bar{\alpha}$as a function of $\bar{\beta}$, we can draw a two-dimensional image of the black hole shadow. The details of the GUP black hole shadow are found by numerical calculation of Model I and Model II (see Figures 1 and 2).

For model I, when the first-order momentum correction coefficients $\alpha$ and second-order momentum correction coefficients $\beta$ of GUP are zero, the black hole shadow shape degenerates to the Kerr black hole case.
When the first-order momentum correction coefficients $\alpha$ and second-order momentum correction coefficients $\beta$ of GUP are both non-zero, the black hole shadow is determined by the model parameters $\alpha$ and $\beta$.
When the black hole spin and second-order momentum correction coefficient $\beta$ are fixed, the black hole shadow scale decreases with the increase of first-order momentum correction coefficient $\alpha$.
When the black hole spin and the first-order momentum correction coefficient $\alpha$ are fixed, the black hole shadow scale increases with the increase of the second-order momentum correction coefficient $\beta$.
That is to say, the first-order momentum correction coefficient $\alpha$ and the second-order momentum correction coefficient $\beta$ change the shadow size of the black hole in the opposite direction.

When the black hole spin $a=0.99$, with the increase of the first-order momentum correction coefficient $\alpha$, the black hole shadow scale is distorted more and more, so there is a certain degree of devolution between the first-order momentum correction coefficient $\alpha$ and the influence of the black hole spin on the black hole shadow shape.
When considering the larger value of the second order momentum correction coefficient $\beta$, the black hole shadow shape is very close to the spherical symmetric black hole case even if the black hole spin is close to $1$.
The significant influence of the first-order momentum correction coefficients $\alpha $and the second-order momentum correction coefficients $\beta$ on the shape of the black hole shadow makes it possible to test it with EHT observations.

For model II, with the increase of the value of the mixing correction coefficient $\varepsilon$, the black hole shadow scale gradually decreases, but its value changes very slowly, which makes it possible to test it with EHT unless the mixing correction coefficient $\varepsilon$ changes sharply.

In fact, since both model I and Model II must satisfy the generalized uncertainty principle (\ref{1-GUP}), the Hawking temperature corresponding to the black hole metric (\ref{2-spacetime elements}) and the black hole metric (\ref{7}) should be consistent, which makes the parameters of model I and model II satisfy the relation (\ref{10}).
Even if the two models can be coordinated by adjusting the values of the parameters, the variation rules of the black hole shadow shape are different because the black hole space-time metric corresponding to model I and model II is different. In view of this, we believe that future EHT observations can be used to test it in order to better understand the quantum gravitational effects of black holes.

\section{Weak cosmic censorship conjecture in GUP-BH and co-constraints}
\label{WCC-test}
In section \ref{shadow} we discuss the possibility of using future EHT observations to examine GUP black holes. Next we are interested in how GUP affects the nature of black holes, such as whether the weak cosmic censorship conjecture of black holes is violated when consider the GUP model. The work and methods of testing the weak cosmic censorship conjecture by particle incident can be referred to a series of literatures e.g.\cite{2019PhRvD.100d4043N, 1974Waldannas, 2017PhRvD..96j4014S, 2018arXiv181203459C}. The work and methods of testing the weak cosmic censorship conjecture by scalar field incident can be referred to a series of literatures e.g.\cite{2015PhRvD..92h4044C, 2021JCAP...10..012G, 2021NuPhB.96515335L, 2013PhRvD..88f4043D, 2018JHEP...09..081G, 2015PhRvD..92j4021S, 2008PhRvL.100l1101H}
Here, we will examine whether the event horizon of the GUP black hole can be destroy by incident the test particle and scalar fields to the GUP black hole, and discuss the influence of GUP on the weak cosmic censorship conjecture.

\subsection{WCCC was tested via particle incidence}
\label{WCC-particle}
According to the discussion of Model I and Model II in Section \ref{Kerr-like}, the analytical expression of the change of GUP model parameters (Model I, model parameters are $\alpha$ and $\beta$; Model II, the model parameter is $\varepsilon$) on the event horizon of the black hole can be obtained, and the basic results are summarized as
\begin{equation}
r_{H}=\left\{ \begin{aligned}
& M\left[1-\frac{2\alpha}{M}+\frac{4\beta}{M^{2}} \right]+ \sqrt{M^{2}\left(1-\frac{2\alpha}{M}+\frac{4\beta}{M^{2}} \right)^{2}-a^{2}} \quad \quad \quad Model I \quad Kerr-like  \\
& M + \sqrt{(1-\varepsilon)M^{2}-a^{2}}, \quad \quad \quad Model II \quad Kerr-Newman-like \\
\end{aligned}\right.
\label{WCCC1}
\end{equation}
It can be found by equation (\ref{WCCC1}). For model I, if the space-time line element (18) describes a black hole, the condition $M^{2}\left(1-\frac{2\alpha}{M}+\frac{4\beta}{M^{2}}\right)^{2}\geq a^{2}$ must be satisfied. For model II, if the space-time line element (19) describes a black hole, the condition $(1-\varepsilon)M^{2}\geq a^{2}$ must be satisfied. The conditions for destroying the event horizon of a black hole are as follows. For model I we need to satisfy the inequality
\begin{equation}
M^{2}\left(1-\frac{2\alpha}{M}+\frac{4\beta}{M^{2}}\right)^{2}<a^{2}.
\label{WCCC2}
\end{equation}
For model II we need to satisfy the inequality 
\begin{equation}
(1-\varepsilon)M^{2}<a^{2}.
\label{WCCC3}
\end{equation}
The inequality equations (51) and (52) can be rewritten as the following expressions
\begin{equation}
J>M^{2}\left(1-\frac{2\alpha}{M}+\frac{4\beta}{M^{2}}\right)
\label{WCCC4}
\end{equation}
and
\begin{equation}
J>M^{2}\sqrt{1-\varepsilon}.
\label{WCCC5}
\end{equation}
The rotating GUP black hole in the equatorial plane incident a test particle with mass, at which time the GUP black hole and the test particle will have a complex interaction. 
When the test particle cross the event horizon of the GUP black hole, both the mass and angular momentum of the GUP black hole increase(Labeled $\delta E$ and $\delta J$). 
By examining whether the mass and angular momentum of the composite system satisfy the event horizon destroying condition (inequality equations (\ref{WCCC4}) or (\ref{WCCC5})), we can analyze how the GUP parameters affect the testing of the weak cosmic censorship conjecture.
The calculation procedure employed here adheres to the standard methodology. We will solely present the outcomes along with their corresponding analysis.

The results show that the event horizon of the GUP black hole can be destroyed when the mass increment ($\delta E$) and angular momentum increment ($\delta J$) of the test particle meet certain conditions. For model I, this condition is
\begin{equation}
\left(1-\frac{2\alpha}{M}+\frac{4\beta}{M^{2}}\right)\delta E^{2}+2\left(1-\frac{2\alpha}{M}+\frac{4\beta}{M^{2}}\right)M\delta E+M^{2}\left(1-\frac{2\alpha}{M}+\frac{4\beta}{M^{2}}\right)-J <\delta J<\frac{1}{\Omega_{H}}\delta E.
\label{WCCC6}
\end{equation}
For model II, the condition becomes
\begin{equation}
\sqrt{1-\varepsilon}\delta E^{2}+2\sqrt{1-\varepsilon}M\delta E+M^{2}\sqrt{1-\varepsilon}-J <\delta J<\frac{1}{\Omega_{H}}\delta E.
\label{WCCC7}
\end{equation}
If the GUP black hole is an extremely rotating black hole, when considering the first approximation of $\delta E$, there is no new term in the black hole event horizon angular velocity $\Omega_{H}$, which makes it possible for no test particle to satisfy the event horizon destroying condition (\ref{WCCC6}) and (\ref{WCCC7}).
This shows that for both Model I and Model II, the introduction of GUP does not violate the weak cosmic censorship conjecture of rotating black holes.
For the near-extreme black hole case, a parameter is introduced to describe the degree of close to the extreme black hole, and then the destroy condition (\ref{WCCC6}) and (\ref{WCCC7}) are simplified. The results show that the black hole event horizon cannot be destroyed for Model I and Model II. The weak cosmic censorship conjecture is not violated.

\subsection{WCCC was tested via scalar field scatter}
\label{WCC-scalar}
In addition to testing the weak cosmic censorship conjecture by incident test particles on a rotating GUP black hole, another method can be used to test the weak cosmic censorship conjecture of a black hole, that is, incident scalar fields on an extreme or near-extreme black hole to destroy the event horizon of the black hole. If the event horizon can be destroyed within the range of optional GUP model parameters, then the weak cosmic censorship conjecture violates. Otherwise, there is no violation.

If there is a scalar field near the GUP rotating black hole, then the scalar field will interact with the black hole, and in the semi-classical case, the equation of motion of the scalar field can be calculated in the black hole space-time background. If the scalar field is labeled with $\psi$ and the mass of the corresponding particle is labeled with $\mu$, then the equation of motion of the scalar field in the space-time background of the GUP rotating black hole satisfies the Klein-Gordon equation, whose basic form is
\begin{equation}
\dfrac{1}{\sqrt{-g}}\partial_{\mu}(\sqrt{-g}g^{\mu\nu}\partial_{\nu}\psi)-\mu^{2}\psi=0,
\label{WCCC20}
\end{equation}
where $g$ is the determinant of the GUP rotating black hole metric, and $g^{\mu\nu}$ is the inverse form of the space-time metric, both of which can be calculated by the space-time metric (\ref{18}) and (\ref{19}). By substituting the GUP black hole metric (\ref{18}) and (\ref{19}) into the equation of motion (\ref{WCCC20}), it can be simplified to the following form
\begin{equation}
-\dfrac{(r^{2}+a^{2})^{2}-a^{2}\Delta\sin^{2}\theta}{\Delta\Sigma^{2}}\dfrac{\partial^{2}\psi}{\partial t^{2}}-\dfrac{4aM}{\Delta\Sigma^{2}}\dfrac{\partial^{2}\psi}{\partial t \partial\phi}+
\dfrac{1}{\Sigma^{2}}\dfrac{\partial}{\partial r}\left( \Delta\dfrac{\partial\psi}{\partial r} \right) 
+\dfrac{1}{\Sigma^{2}\sin\theta}\dfrac{\partial}{\partial \theta}\left( \sin\theta\dfrac{\partial\psi}{\partial \theta} \right) 
+\dfrac{\Delta-a^{2}\sin^{2}\theta}{\Delta\Sigma^{2}\sin\theta}\dfrac{\partial^{2}\psi}{\partial\phi^{2}}-\mu^{2}\psi=0.
\label{WCCC21}
\end{equation}
This equation can be processed by variable separation, which decomposes $\psi$ into $\psi(t,r,\theta,\phi)=e^{-i\omega t}R(r)S_{lm}(\theta)e^{im\phi}$, where $S_{lm}(\theta)$ is a spherical function, and the integers $l$ and $m$ are quantum numbers. 
By separating variables in this way, equation (\ref{WCCC21}) can be simplified into two equations, one of which is an angular equation, the solution of which is a spherical harmonic function, and the other is a radial equation, whose specific expression is
\begin{equation}
\dfrac{d}{dr}\left( \Delta\dfrac{dR}{dr} \right)+
\left[ \dfrac{(r^{2}+a^{2})^{2}}{\Delta}\omega^{2}-\dfrac{4aM}{\Delta}m\omega+\dfrac{m^{2}a^{2}}{\Delta}-\mu^{2}r^{2}-\lambda_{lm} \right]R(r)=0.
\label{WCCC22}
\end{equation}
By examining equation (\ref{WCCC22}), we can find that if we want to obtain an analytic expression of the radial function $R(r)$ near the event horizon, since $\Delta=0$ near the event horizon, equation (\ref{WCCC22}) will lose its physical meaning (this is because equation (\ref{WCCC22}) corresponds to the BL coordinate system).
Therefore, the equation (\ref{WCCC22}) needs to be converted to other coordinate systems in order to solve it, in general, the Eddington-Finkelstein coordinate system (also known as turtle coordinate) is selected, and for radial coordinates, the following relationship is satisfied
\begin{equation}
\dfrac{dr}{dr_{*}}=\dfrac{\Delta}{r^{2}+a^{2}}.
\label{WCCC23}
\end{equation}
The turtle coordinate $r_{*}$ at the GUP rotating black hole event horizon does not make equation (\ref{WCCC22}) singular. Substituting the coordinate transform (\ref{WCCC23}) into (\ref{WCCC22}) transforms the radial equation (\ref{WCCC22}) into
\begin{equation}
\dfrac{\Delta}{(r^{2}+a^{2})^{2}}\dfrac{d}{dr}(r^{2}+a^{2})\dfrac{dR}{dr_{*}}+\dfrac{d^{2}R}{dr_{*}^{2}}+\left( \omega-\dfrac{ma}{r^{2}+a^{2}} \right)^{2}R(r)
+\left[\dfrac{\Delta}{(r^{2}+a^{2})^{2}}2am\omega-\dfrac{\Delta}{(r^{2}+a^{2})^{2}}(\mu^{2}r^{2}+\lambda_{lm})\right]R(r)=0.
\label{WCCC24}
\end{equation}
Near the event horizon of a GUP rotating black hole(for Model I and Model II), $\Delta=0$ simplifies the equation (\ref{WCCC24}) to the following equation
\begin{equation}
\dfrac{d^{2}R}{dr_{*}^{2}}+( \omega- m\Omega_{H})^{2}R=0.
\label{WCCC25}
\end{equation}
This equation is a second order ordinary differential equation, and its analytical solution is
$R(r)=\exp[\pm i(\omega-m\Omega_{H})r_{*}]$. Plug it into the variable separation expression of $\psi$, whose final solution is
\begin{equation}
\psi(t,r,\theta,\phi)=\exp[- i(\omega-m\Omega_{H})r_{*}]e^{-i\omega t}S_{lm}(\theta)e^{im\phi}.
\label{WCCC26}
\end{equation}
With this solution, we have obtained all the information of the scalar field near the event horizon, and can use it to calculate various physical quantities.

Suppose that a scalar field is incident into the GUP rotating black hole space-time background. Due to the maximum effective potential of the black hole space-time in the radial direction, the space-time will absorb and reflect the scalar field. These physical processes change the energy and angular momentum of the black hole.
The corresponding energy increment and angular momentum increment can be calculated from the scalar field (\ref{WCCC26}). For the scalar field considered in this section, the expression of the energy-momentum tensor can be obtained for the scalar field moving in the space-time background of a GUP rotating black hole
\begin{equation}
T_{\mu\nu}=\partial_{\mu}\psi \partial_{\nu}\psi^{*}-\dfrac{1}{2}g_{\mu\nu}(\partial_{\mu}\psi \partial^{\nu}\psi^{*}+\mu^{2}\psi^{*}\psi).
\label{WCCC27}
\end{equation}
By combining the expression of the energy momentum tensor (\ref{WCCC27}) with the scalar field (\ref{WCCC26}), all the energy- momentum non-zero components of the scalar field can be obtained. Using these non-zero components, we can calculate the energy flow and angular momentum flux of the scalar field incident on the black hole. The energy flow of a scalar field through the event horizon is
\begin{equation}
\dfrac{dE}{dt}=\int\limits_{H} T_{\quad t}^{r}\sqrt{-g}d\theta d\phi=\omega(\omega-m\Omega_{H})(r_{H}^{2}+a^{2}),
\label{WCCC28}
\end{equation}
the angular momentum flux of a scalar field through the event horizon is
\begin{equation}
\dfrac{dJ}{dt}=\int\limits_{H} T_{\quad \phi}^{r}\sqrt{-g}d\theta d\phi=m(\omega-m\Omega_{H})(r_{H}^{2}+a^{2}).
\label{WCCC29}
\end{equation}
From these two expressions, it can be seen that whether the incident scalar field transfers energy or angular momentum to the black hole depends on the relative size of $\omega$ and $m\Omega_{H}$. In a very short time, the energy increment and angular momentum increment of the scalar field transferred to the black hole are
\begin{equation}
dE=\omega(\omega-m\Omega_{H})(r_{H}^{2}+a^{2})dt
\label{WCCC30}
\end{equation}
and
\begin{equation}
dJ=m(\omega-m\Omega_{H})(r_{H}^{2}+a^{2})dt.
\label{WCCC31}
\end{equation}
For a scalar field incident on a black hole, it can be divided into an infinite number of small segments ($dt$) for segmentation discussion. Assume that the mass and angular momentum of the GUP rotating black hole before the scalar field incident on the black hole are $M$ and $J$, respectively. The mass and angular momentum of the GUP rotating black hole-scalar field complex system are $M^{'}$ and $J^{'}$, respectively. They satisfy the relation $M^{'}=M+dE$ and $J^{'}=J+dJ$.
Then, by examining the signs of $M^{'2}\omega_{0}-J^{'}$ of the composite system, we can determine whether the event horizon of the composite system exists, and whether the weak cosmic supervision conjecture is violated when the scalar field incident on the GUP rotating black hole. Next, we will only give the calculation results, and the calculation process can be referred to XX.

By calculation, for a scalar field of mode $(l,m)$, the expression $M^{'2}\omega_{0}-J^{'}$ for a composite system is
\begin{equation}
M^{'2}\omega_{0}-J^{'}
=(M^{2}\omega_{0}-J)+2M\omega_{0}m^{2}\left( \dfrac{\omega}{m}-\dfrac{1}{2M\omega_{0}} \right)\left( \dfrac{\omega}{m}-\Omega_{H} \right)(r_{H}^{2}+a^{2})dt.
\label{WCCC32}
\end{equation}
As long as we can judge the sign of (\ref{WCCC32}), we can discuss whether the scalar field can destroy the event horizon. When considering extreme black holes, since extreme black holes satisfy the condition $J=M^{2}\omega_{0}$, then $M^{'2}\omega_{0}-J^{'}$ is reduced to the following form
\begin{equation}
M^{'2}\omega_{0}-J^{'}
=2M\omega_{0}m^{2}\left( \dfrac{\omega}{m}-\dfrac{1}{2M\omega_{0}} \right)\left( \dfrac{\omega}{m}-\Omega_{H} \right)(r_{H}^{2}+a^{2})dt.
\label{WCCC33}
\end{equation}
For model I and Model II, for extreme black holes, since $\Omega_{H}$ does not have any shift, that is, $\Omega_{H}=\dfrac{1}{2M\omega_{0}}$, then $M^{'2}\omega_{0}-J^{'}$ is always greater than zero, which indicates that for a scalar field incident extreme GUP black hole, the event horizon of the black hole cannot be destroy, and the weak cosmic censorship conjecture is not violated.
For the case of near-extreme GUP black holes, a parameter describing the degree of near-extreme black holes can also be introduced for discussion (similar to the analysis techniques in Section \ref{WCC-particle}). The calculation results show that $M^{'2}\omega_{0}-J^{'}$ is also always greater than zero for near-extreme black holes, so the black hole event horizon cannot be destroy, and the weak cosmic censorship conjecture is still not violated.

\section{Summary}
\label{summary}

In this study, we obtain the rotating black hole solutions of the Einstein gravitational field equations when the GUP modifications are taken into account and discuss the physical properties of these solutions. On the physical picture, we construct the GUP modifications to the Kerr black hole in two different ways. Model $I$ mainly considers GUP only for mass modification, which changes our understanding of gravitational mass and inertial mass, and this construction is closer to physical nature. Model $II$ considers the GUP as an effect of an external field, so the energy-momentum tensor corresponding to the black hole system is not zero. In this case, the physical meaning of GUP for black hole modification is completely different from that of Model $I$. From the perspective of black hole thermodynamics. For spherical symmetry GUP black hole case, Model $I$ and Model $II$ should be equivalent, so we get the relationship between model parameters $\alpha$, $\beta$ and $\varepsilon$ (\ref{10}). Although the GUP black hole modifications constructed by these two methods are consistent in black hole thermodynamics, their space-time properties are different. For the GUP spinning black hole case, the thermodynamics are not equivalent (the Hawking temperature corresponding to Model I and Model II are not equal). The black hole metric $(\ref{18})$ obtained by Model $I$ is similar to Kerr black hole, and the black hole metric $(\ref{19})$ obtained by Model $II$ is similar to Kerr-Newman black hole. 

The potential of black hole shadows in testing GUP is explored based on the spacetime of a rotating black hole (18) and (19), leading to the following findings. 
On one hand, when both the first-order momentum correction coefficient $\alpha$ and the second-order momentum correction coefficient $\beta$ of GUP are non-zero, the determination of the black hole shadow relies on the model parameters $\alpha$ and $\beta$. Furthermore, these coefficients exhibit an opposite effect on altering the size of the black hole shadow, which is a significant characteristic that may potentially be confirmed in future EHT observations.
On the other hand, when considering the higher magnitude of the second order momentum correction coefficient $\beta$, even in cases where the black hole spin approaches $1$, the shape of the black hole shadow closely resembles that of a spherically symmetric black hole. This observation suggests a potential degenerate effect between the second order momentum correction coefficient $\beta$ and the black hole spin.

The weak cosmic censorship conjecture for GUP rotating black holes is also investigated simultaneously. By introducing test particles and scalar fields into the GUP rotating black hole, we find no violation of the weak cosmic censorship conjecture in both extreme and near-extreme black hole scenarios. Hence, even with the modification of black hole spacetime due to GUP, it does not alter the discourse on the nature of black holes.
The weak cosmic censorship conjecture remains unaffected by GUP, however, it cannot be guaranteed that the strong cosmic censorship conjecture will not be influenced. In future research, we will investigate the applicability of the strong cosmic censorship conjecture to GUP rotating black holes.

\begin{acknowledgments}
We acknowledge the anonymous referee for a constructive report that has significantly improved this paper. This work was supported by the  Special Natural Science Fund of Guizhou University (Grants
No. X2020068, No. X2022133) and the National Natural Science Foundation of China (Grant No. 12365008).
\end{acknowledgments}

\end{document}